# Trends in students media usage

Gerd Gidion[1], Luiz Fernando Capretz[2], Michael Grosch[1] and Ken N. Meadows[2]

[1]Karlsruhe Institute of Technology, Karlsruhe, Germany
`gidion@kit.edu, michael.grosch@kit.edu`
[2]Western University, London, Ontario, Canada
`lcapretz@uwoa.ca, kmeadow2@uwo.ca`

**Abstract.** Trends in media usage by students can affect the way they learn. Students demand the use of technology, thus institutions and instructors should meet students' requests. This paper describes the results of a survey where drivers in the use of media show continuously increasing or decreasing values from the first to the fourth year of study experience at the Western University, Canada, highlighting trends in the usage of new and traditional media in higher education by students. The survey was used to gather data on students' media usage habits and user satisfaction from first to fourth year of study and found that media usage increases over the years from first to fourth. The presentation of data using bar charts reveals a slight increase over the years in students owning notebooks or laptops off-campus and a significant increase from first to fourth year of students accessing online academic periodicals and journals. Another noteworthy finding relates to fourth year students being more conscious of the quality of information that they read on the Internet in comparison to students in first year, even though this is a slight year on year increase.

## 1 Introduction

Digital media has changed students' learning environments and behaviors in higher education (Venkatesh et al., 2014). The NMC Horizon Report (Johnson et al., 2014) shows that a rapid and ubiquitous diffusion of digital media into higher education has led to changes in the students' learning environments and it has also influenced their learning behaviors. To keep pace with this changing learning environment, post-secondary institutions need to understand and analyze media usage behaviors of their students. This study, focusing on the media usage habits of students, was designed to provide an evidence base upon which reliable predictions can be made about future trends in media usage in higher education. The framework on which the research is based posits that the current teaching and learning methods are utilizing, and are influenced by, media that are a combination of traditional (e.g., printed books and journals) and new media (e.g., Wikipedia, Google). The current state of teaching and learning has been influenced by former media usage habits and these habits tend to change with the introduction of new media. In the future, teaching and learning will be influenced by the rising new media usage habits as well as by the current teaching and learning paradigms

A second focus of this research is on media acceptance which provides an indicator of media quality from the participants' perspectives. Therefore, media quality is assessed by measuring the acceptance of the services used by students.

The Media Usage Survey was developed to provide researchers with a deeper and more detailed understanding of students' technology usage in learning and of possible environmental factors that may influence that usage.

## 2      Literature review

Students tend to be early adopters of media and information technology, as they possess ample opportunities to access media, encouraged by their curiosity and self-learned skills. But they are not just passive users of technology, they are the designers and developers of the technology as well. For example, Google, the most commonly used search engine on the Internet, was created by Stanford students in the late 90s and Facebook was created by Harvard University students in 2004 and in less than ten years became one of the most successful Internet services worldwide.

Students in post-secondary education intensively use web services, such as Google, Wikipedia, and Facebook during their free time as well as for their studies (Smith et al. 2009). Current development in the so-called web 2.0 is often characterized by the increase in interactions between users (Capuruco & Capretz, 2009). This is even more apparent with the more recent rise of collaborative media - in the form of social networking sites, file-sharing, wikis, blogs, and other forms of social network software.

Buckingham (2007) asserts that one cannot teach about the contemporary media without taking into account the role of the Internet, computer games, and the convergence between 'old' and 'new' media. Much of the popular discussion in this area tends to assume that students already know everything about the new media; they are celebrated as 'millenials' (Howe & Strauss, 2000), or as "digital natives" (Prensky, 2001) who are somehow spontaneously competent and empowered in their dealings with new media, although other researchers argue about lack of evidence (Bennett et al, 2008; Bennett and Maton, 2012; Helsper & Enyon, 2009). They learn to use these media largely through trial and error – through exploration, collaboration with others, experimentation and play.

Pritchett et al. (2013) examined "the degree of perceived importance of interactive technology applications among various groups of certified educators" (p. 34) and found that, in the involved schools, some groups seem to perceive Web 2.0 media as more important than others, e.g. participants of the survey with "an advanced degree and/or higher certification level" (p. 37).

Furthermore, mobile broadband Internet access and the use of corresponding devices, such as netbooks and smartphones, have fueled the boom of the social networks by students in higher education. Murphy et al. (2013) reported that in spite of the limitations in the formal university infrastructure, many students would like to use their mobile devices for formal learning as well as informal learning. Recent development in technology, with smartphones and tablets dominating the market in recent years have ensured that these devices have great functionality and enable interactivity, thus fulfilling the desire for both formal learning (Lockyer & Patterson, 2008) and informal learning (Johnson & Johnson, 2010).

There has been suggestion concerning the potential of this technological shift in students' learning and the real benefits of these technologies for learning (Johnson et al. 2014). There has been considerable research demonstrating the costs and benefits of using social, mobile, and digital technology to enhance teaching and learning; yet the research is not conclusive as to whether the use of these technologies leads to improved learning outcomes (Cusumano, 2013;). Klassen (2012) states: "If there is one thing I have learned the last ten years about the use of new technology in education, it is that the combination of old and new methods make for the best model."; and goes on to say: "Students will continue to seek out inspiring teachers. Technology alone is unlikely to ensure this, although it may make a lot of average teachers seem a lot better than they are!"

The usage of media at a university is a topic of interest for students, staff and faculty. There may be diverse interests and habits, but several interdependencies and interactions. The understanding of one of these scenarios has been the objective of a study by Kazley et al. (2013). They surveyed these groups and defined certain "factors that determine the level of educational technology use" (p. 68). They describe a model with increasing intensity / quality of technology use, from beginners (using email and basic office software) to experts (using videoconference, virtual simulation tools etc.).

There is no doubt, however, that the integration of IT media and services in higher education appears to have led to substantial changes in the ways in which students study and learn (Dahlstrom, 2012). Higher education institutions are cautious about investing in programs to provide students with mobile devices for learning, due to the rapidly changing nature of technologies (Alrasheedi & Capretz, 2013a). The acceptance of technology-enhanced education by students has increased in recent years, but not all services are equally accepted (Alrasheedi & Capretz, 2013b). It has become clear that simply using media and adopting e-learning does not necessarily make a difference in student learning. Rather, key factors for effective use of technology are pedagogy and the quality of the services (Alrasheedi & Capretz, 2013c).

Moreover, we contend that, despite the highly contextual nature of e-learning studies, several characteristics are similar and the results could be developed into a framework for the assessment of the success of e-Learning. One such framework was presented by Ali, Ouda & Capretz (2012), where learning contexts, learning experience, and design aspects were used to assess the success of mobile learning. Also, Capuruco & Capretz (2009) developed a social-aware framework for interactive learning.

In summary, the variety of media enriched informal learning processes is relevant. This perspective on the whole spectrum of media used for learning (printed, e-learning, digital, web 2.0, etc.) requires a certain theory-oriented empirical research approach to reach a deeper understanding about the media usage behavior of student in higher education.

## 3    Methods and procedures

The integration of IT media and services in higher education has led to substantial changes in the ways in which students study and learn, and instructors teach (Johnson et. al., 2014). This survey was designed to measure the extent to which media services are used in teaching and learning as well as to assess changes in media usage patterns. The survey is a landmark, as it is the first of its kind in Canada and represents an initial foray into the North American post-secondary sector.

The survey of students' media usage habits was conducted at Western University in London, Ontario, Canada, in 2013. The survey focuses primarily on the media usage habits of students. Based on an assessment of the way in which media use relates to teaching and learning, the identification of trends provides an evidence base upon which more reliable predictions can be made about future trends of media usage in higher education.

The survey is anonymous and consists of 150 items measuring frequency of media usage and user satisfaction with 53 media services, including:

- Media hardware such as Wi-Fi, notebooks, tablet computers, desktop computers, and smartphones;
- Information services, such as Google search, Google Books, library catalogues, printed books, e-books, printed journals, e-journals, Wikipedia, open educational resources, and bibliographic software;
- Communication services, such as internal and external e-mail, Twitter, and Facebook;
- e-learning services and applications, such as learning platforms and wikis.

Additional variables were also evaluated such as some aspects of learning behavior, media usage in leisure time, educational biography, and socio-demographic factors, etc.

The survey tool was first developed in 2009 and used at Karlsruhe Institute of Technology (KIT) in Germany (Grosch & Gidion, 2011). During the course of 15 follow-up surveys that were administered in a variety of countries, the original survey underwent customization, translation into several languages, and validation.

In this study, the survey was administered at The University of Western Ontario in London, Ontario, Canada to undergraduate students in the Winter of 2013 academic term. Having data from students from first to fourth years allows the researchers to compare their media usage to gain insight into possible discrepancies in their respective media cultures. These discrepancies could conceivably result in problems in the use of media for studying and teaching. The data for this survey was collected online using a well-established online survey tool: Unipark.

## 4 Results

Data collection occurred in two waves. In the first wave, 300 students in their first year in Engineering were surveyed in the Fall of 2012 (ESt1styear2012) but only 100 students completed the survey. Partial results involving instructors and students only in the Faculty of Engineering were presented at the Canadian Engineering Education Association Conference (Gidion et al. 2013). A students satisfaction with media survey was also conducted and results are reported in (Gidion et al. 2014).

In the second wave of data collection, conducted between January 16th and February 15$^{th}$, 2013, 19,978 undergraduate students (with 1 year, 2 years, 3 years, 4 and more years of study) from across the faculties were invited to respond to the survey and 1266 participated. Eight hundred and three students completed the survey (i.e., had a completion rate of more than 90% of the survey items). While participants were from a broad spectrum of demographic characteristics and faculties, female students were disproportionately represented as in common in survey research (Sax et al. 2003). Otherwise, with some caveats, respondents are generally representative of the winter 2013 student population at Western.

### 4.1 Media usage of students by academic years

The results show that students most often attend class, followed by studying using a computer, and studying by themselves at home. Searching on the Internet for learning materials seems to be slightly more common than visiting libraries. Cooperative learning, compared to the other habits, seems relatively rare. The influence of cohort or study experience on the general habits to utilize media could have been interpreted by continuously developing usage frequencies. One of the few items in the complete questionnaire that shows (with the exception of the "beginners" from Engineering) a continuous increase is "study using a computer", but this happens anyway on a high level for all subgroups. This trend is presented in Figure 1.

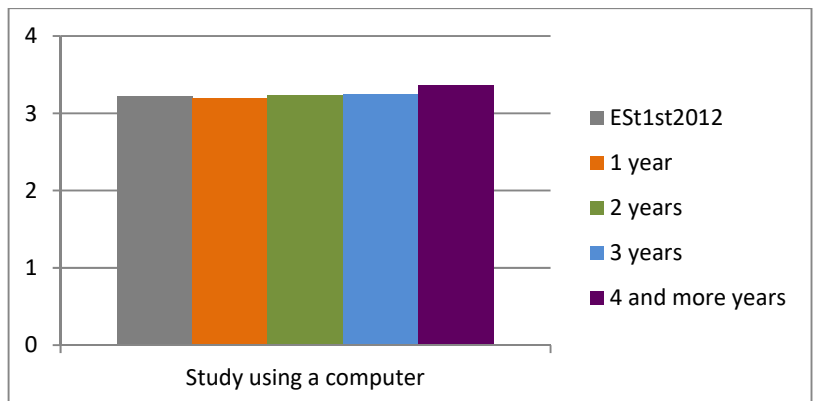

Fig. 1. Respondents mean use of a computer for studying, rated from "never" (0) to "very often" (4).

### 4.2 Increasing media use

For a subset of media, there is an increase in their use across the four years of the degree. These media includes respondents' notebook/laptop off campus, printed books and printed handouts from instructors, Wikipedia, and Google books as well as online services from the universities library and bibliographic software. Interestingly, there are large increases over the course of the four years in the use of "e-versions of academic periodicals/journals", an increase the strength of which is not evident with printed versions of these journals. An Analysis of Variance (ANOVA) was performed to examine these results. Significant differences between the groups (program year) could be found concerning the usage frequency of own notebook / laptop off-campus, Computer labs on campus (e.g. Genlab), online services of the university library (central) / faculty library, e-versions of academic periodicals / journals and Wikipedia. Computer labs on campus are increasingly used in higher terms. This trend is depicted in Figure 2.

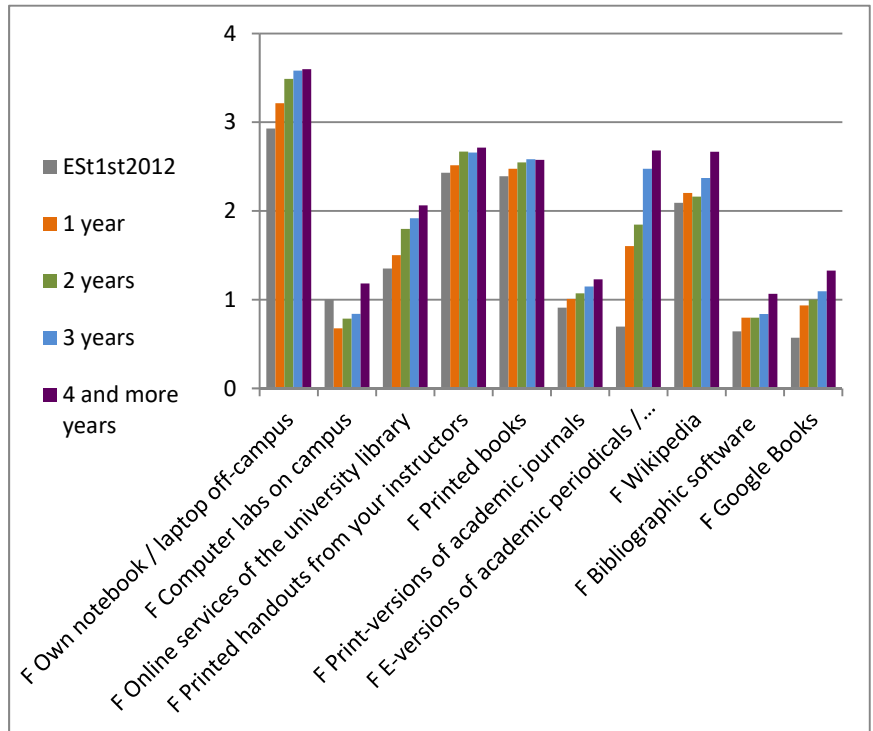

Fig. 2. Mean frequency of respondents use of different media for learning by year of program, rated from "never" (0) to "very often" (4). ['F' stands for Frequency]\

### 4.3 Decreasing media use

Looking at the decreasing media use, the two main trends concern online self-tests for studying and online exams for course grades. The ANOVA showed a significant result concerning the differences between the groups (year of program) for online exams (for grades in a course) and online self-tests with the frequency of use decreasing along the years. This decrease may be the result of the nature of learning assessments across the four years in some faculties, with assessments in the early years lending themselves more to self-testing (e.g., multiple choice examinations) than those often used in the later years (e.g., essay assignments and exams). There was not a significant effect for game-based learning applications. Facebook has a slightly (but significant) shrinking frequency, and on a lower level of frequency the items game-based learning applications, Google+ and virtual class in real-time as well as in non-real-time. This overall tendency is described in Figure 3.

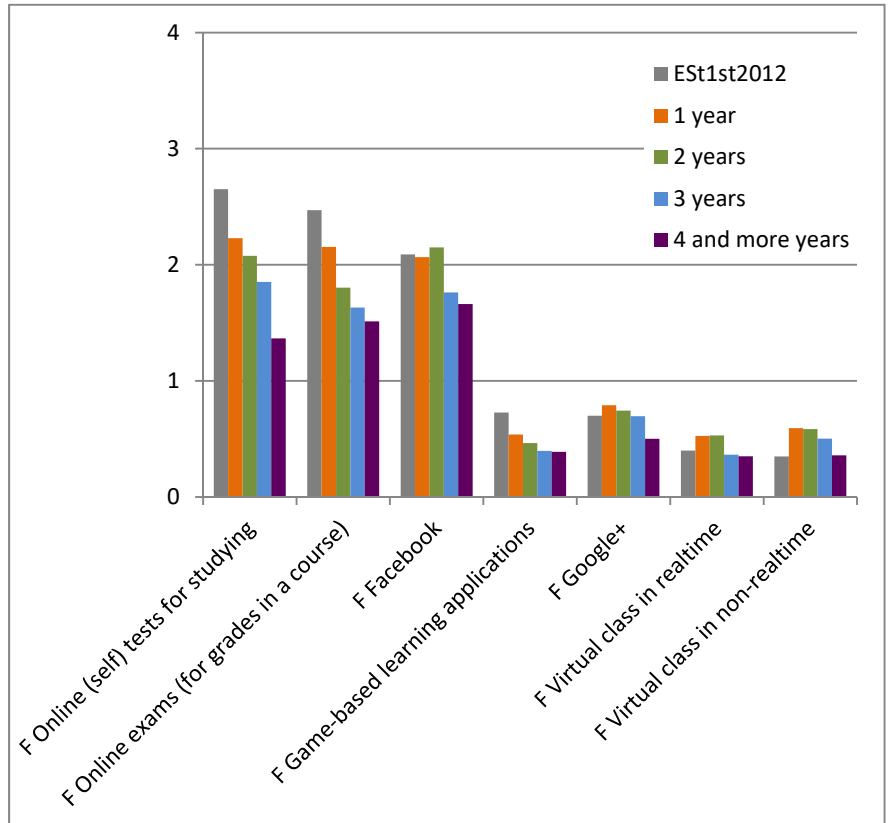

Fig. 3. Mean frequency of respondents use of different media for learning by year of program, rated from "never" (0) to "very often" (4). ['F' stands for Frequency]

### 4.4 Increasing satisfaction with media usage

There were no clear trends which suggest that students' satisfaction with their media use for learning increased across the four years. For the items "own notebook / laptop off campus" and "online dictionary", there are inconsistent increases across the time period. With "game-based learning applications" and "augmented reality applications" students from the first two years of study are less satisfied than students with three and more years at the university (see Figure 4).

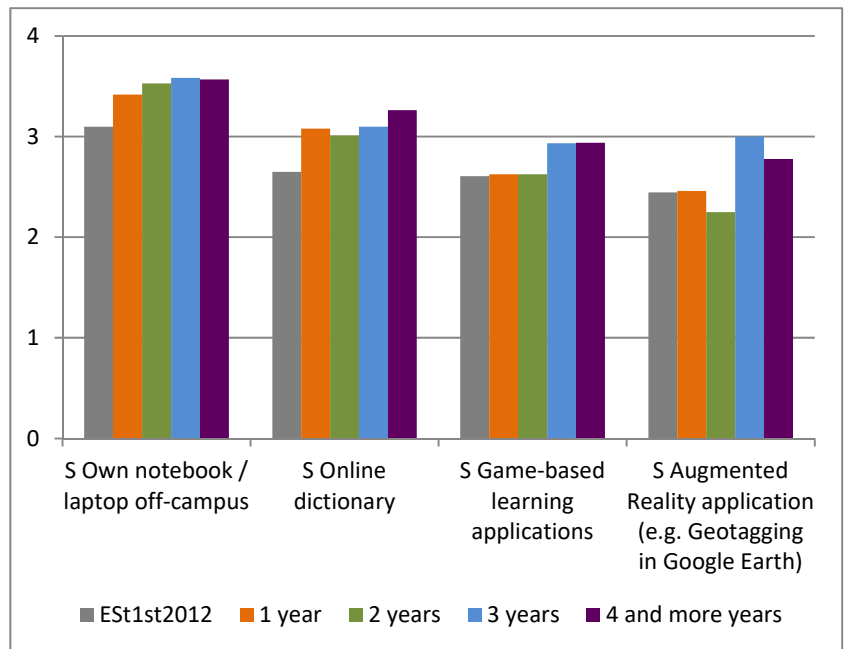

Fig. 4. Mean frequency of respondents' satisfaction with the use of different media for learning by year of program, rated from "never" (0) to "very often" (4). ['S' stands for Satisfaction]

### 4.5 Decreasing appreciation regarding media usage

Figure 5 captures the answer to two questions reflecting the openness of instructors to the use of new media for students' and the instructors own teaching. There is a decreasing trend in terms of the instructors' performed use (significant) and their openness to new media from year to year – following the students perception. It is not clear if this is a result of experience with the finished terms or the changing of openness from year to year of study (e.g. that instructors are not as open for third year students media use compared with the first year students.

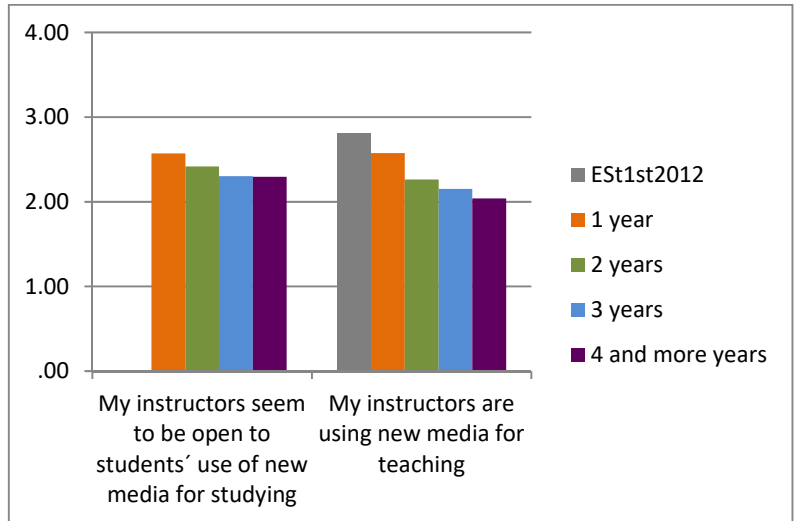

Fig. 5. Means of students' responses to specifically selected items to the question: To what extent do you agree /disagree with the following statements? [0 = strongly disagree, 4 = strongly agree]

**4.6 Increasing information literacy with media usage**

In the group of items with statements the results for students just came to one item with continuous increasing values: to check the quality of information during an internet search seems to be more relevant from year to year at the university (Figure 6).

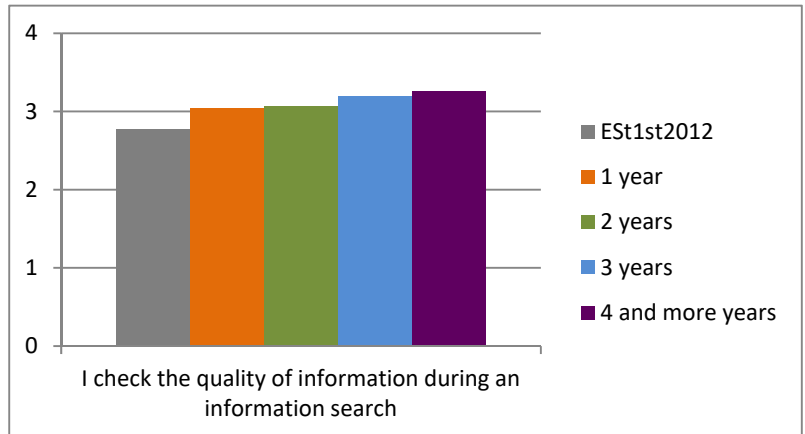

Fig. 6. Means of students' responses to a specifically selected item to the question: To what extent do you agree /disagree with the following statements? [0 = strongly disagree, 4 = strongly agree]

### 4.7 Decreasing satisfaction with media usage

Several media showed a decreasing level of satisfaction from the first to the fourth year, specifically with mobile phones, own notebook/laptop on campus and wireless connection on campus (significant). With the exception of the first year students from Engineering, the respondents rated the universities website, the universities e-mail-account (significant), the Learning Management System (significant), online exams and video sharing websites (significant) as less satisfying from year to year (see Figure 7). Again it is not possible to determine from these data if this decrease is the result of dissatisfaction because of the respondents' experience with the media or if the usage habits are different which is related to the respondents' satisfaction.

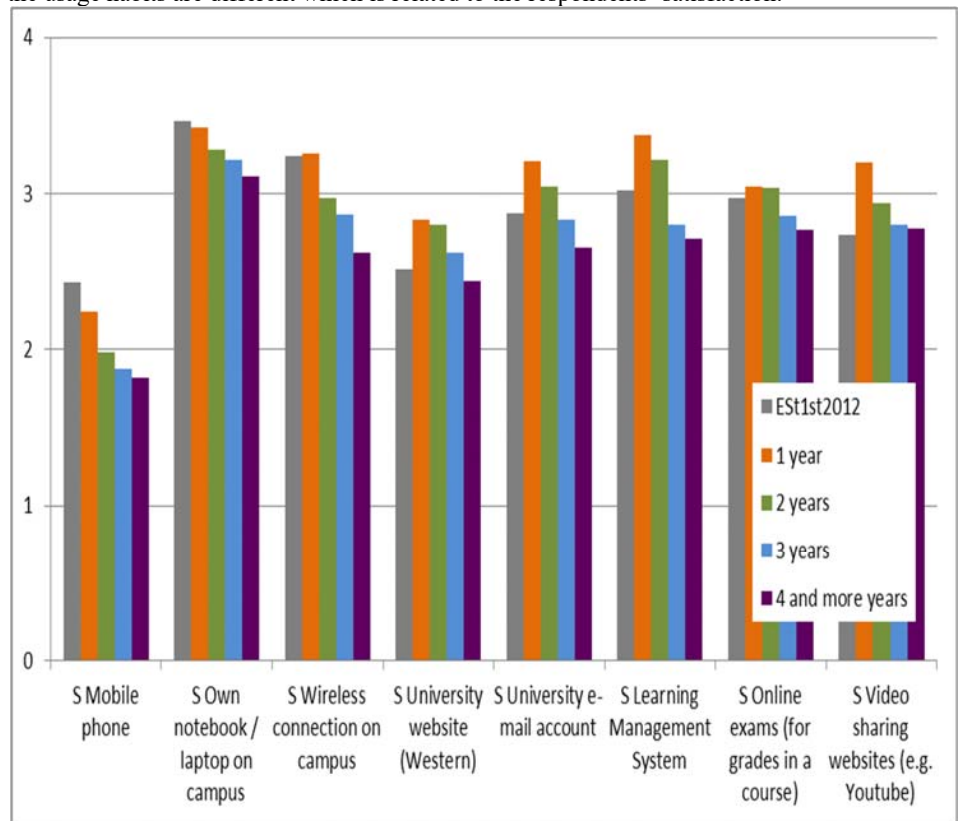

Fig. 7. Mean frequency of respondents' satisfaction with the use of different media for learning by year of program, rated from "never" (0) to "very often" (4). ['S' stands for Satisfaction]

## 5       Conclusion

We examined media usage trends among students in higher education to provide an evidence base so that predictions can be made about future trends regarding students and their media usage and habits. Whilst current teaching and learning combines both old and new media, this study acknowledges the need for an analysis of the current changes in studying and learning based on the integration of IT and new media into higher education.

There is a great deal of value and potential in this research insofar as it presents current data on students' media usage and satisfaction in higher education. Overall the study demonstrates a trend of increased media usage by students. The use of technology increases from $1^{st}$ to $4^{th}$ year students, whereas satisfaction with technology decreases among $1^{st}$ to 4 year student. However, first year student seem to be using social media more frequently than students in other years, whereas 4th year student appear to be more conscious about the quality of information they read on the Internet. A trend of special interest in the next few years will be mobile learning and the impact of BYOD (Bring Your Own Device) in the students' learning environment.